\documentclass{PoS}

\usepackage{amsmath}

\title{Centre vortices are the seeds of dynamical chiral symmetry breaking}

\ShortTitle{Centre vortices are the seeds of dynamical chiral symmetry breaking}

\author{\speaker{Waseem Kamleh}%
\thanks{This research was undertaken with the assistance of resources
  at the NCI National Facility in Canberra, Australia, and the iVEC
  facilities at the University of Western Australia (iVEC@UWA),
  provided through the National Computational Merit Allocation Scheme
  and the University of Adelaide Partner Share. This research is supported by the
  Australian Research Council.} \\
        Special Research Centre for the Subatomic Structure of Matter,\\
        University of Adelaide, Australia\\ 
        E-mail: \email{waseem.kamleh@adelaide.edu.au}}

\author{Derek B. Leinweber \\
       Special Research Centre for the Subatomic Structure of  Matter, \\
       University of Adelaide, Australia}

\author{Daniel Trewartha \\
       Special Research Centre for the Subatomic Structure of  Matter, \\
       University of Adelaide, Australia}

\abstract{ Using lattice QCD, we reveal a fundamental connection
  between centre vortices and several key features associated with
  dynamical chiral symmetry breaking and quark
  confinement. Calculations are performed in pure $\mathrm{SU}(3)$
  gauge theory using the chiral overlap fermion action. Starting from
  the original Monte Carlo gauge fields, a vortex identification procedure yields
  vortex-removed and vortex-only backgrounds. We examine the static
  quark potential, the quark mass function, the hadron spectrum, the
  local topological charge density, and the distribution of
  instanton-like objects on the original, vortex-removed and
  vortex-only ensembles. The removal of vortices consistently results
  in the removal of the corresponding feature associated with
  dynamical chiral symmetry breaking. Remarkably, we observe that
  after some smoothing, in each of these cases, the vortex-only
  degrees of freedom are able to encapsulate the pertinent features
  of the original gauge fields.}

\FullConference{34th annual International Symposium on Lattice Field Theory\\
		24-30 July 2016\\
		University of Southampton, UK}

\begin{document}


Two key features of quantum chromodynamics (QCD) are the confinement
of quarks inside hadrons, and dynamical chiral symmetry breaking,
associated with the dynamical generation of mass. An analytic proof of
the underlying mechanisms responsible for these phenomena remains
elusive. Dynamical chiral symmetry breaking and confinement appear to
be emergent properties of QCD, generally accepted to originate from
the topological structure of the nontrivial vacuum.

The centre vortex model of confinement \cite{'tHooft:1977hy,'tHooft:1979uj} 
is well-known. A centre vortex intersects with a two-dimensional region $A$ of the gauge manifold $U$ if the Wilson loop identified with the boundary has a nontrivial transformation property $U(\partial A) \rightarrow zU(\partial A),\ z \neq 1,$ under an element $Z = zI \in \mathbb{Z}_3$ of the centre group of SU(3), 
where $z \in \{ 1, e^{\pm2\pi i/3} \}$ is a cube root of unity. On the lattice we seek to decompose the gauge links $U_\mu(x)$ in the form
\begin{equation}
U_{\mu}(x) = Z_{\mu}(x)\cdot R_{\mu}(x),
\end{equation}
in such a way that all vortex information is captured in the field of centre-projected elements $Z_{\mu}(x)$, with the remaining short-range fluctuations described by the vortex-removed field $R_{\mu}(x).$ By fixing to Maximal Centre Gauge and then projecting the gauge-fixed links to the nearest centre element, we may identify the vortex matter by searching for plaquettes with a nontrivial centre flux around the boundary. These are the ``thin'' centre vortices (or P-vortices), that are embedded within the ``thick'' centre vortices of the original, Monte Carlo generated, configurations. In this way we create three distinct ensembles of SU(3) gauge fields:
\begin{enumerate}
\item The original `untouched' configurations, $U_{\mu}(x),$

\item The projected vortex-only configurations, $Z_{\mu}(x),$

\item The vortex-removed configurations, $R_\mu(x) = Z^{\dagger}_{\mu}(x)\,U_{\mu}(x).$
\end{enumerate}
Performing lattice calculations of key quantities on the three different ensembles and comparing the differences that emerge from the absence or presence of centre vortices has proven to be a rich source of information regarding the properties of pure Yang-Mills theory. Here we review the recent work that has been performed by the CSSM lattice collaboration~\cite{OMalley:2011aa,Trewartha:2015ida,Trewartha:2015nna}, and direct the reader therein for the details of the results presented here and a comprehensive list of references.

We begin by examining the three ensembles at the microscopic scale through visualisations of the topological charge density. It is well established that some form of cooling or smoothing is required in order to reveal the instanton-like topological structures present in the QCD vacuum. Cooling is a process by which the short-range fluctuations present in the gauge field are smoothed away to reveal the long range physics. Indeed, without any cooling there is little or no visible difference in the topological charge density on the three ensembles. However, the application of cooling reveals some very interesting physics. 

\begin{figure}[t]
  \centering
  \includegraphics[width=0.48\textwidth]{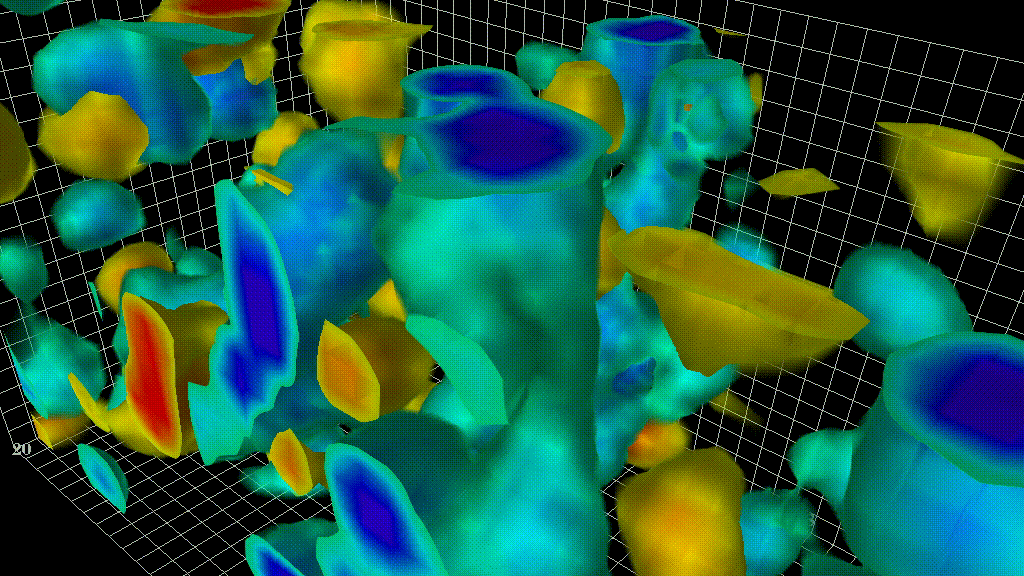}
  \includegraphics[width=0.48\textwidth]{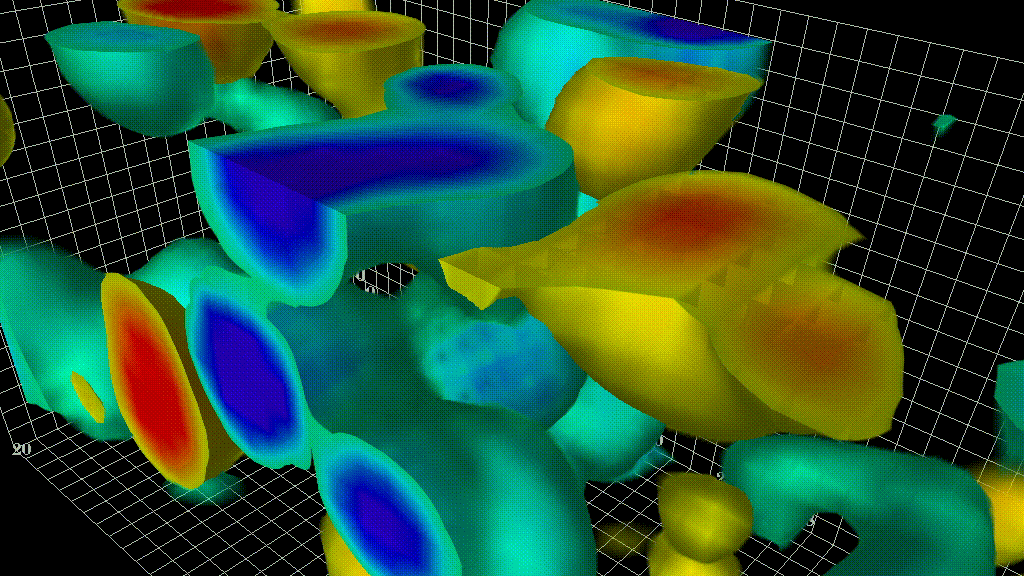}\\
  \includegraphics[width=0.48\textwidth]{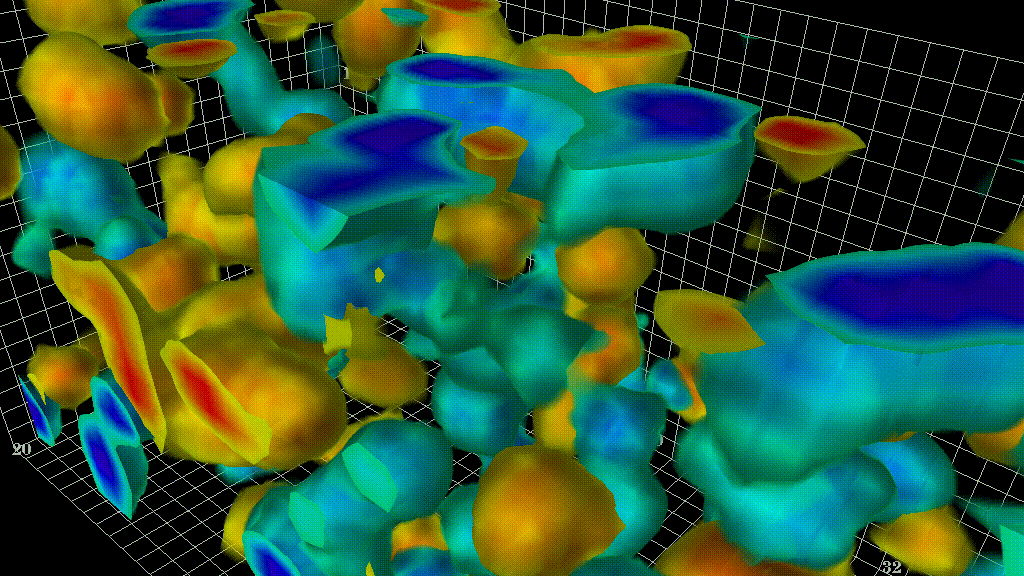}
  \includegraphics[width=0.48\textwidth]{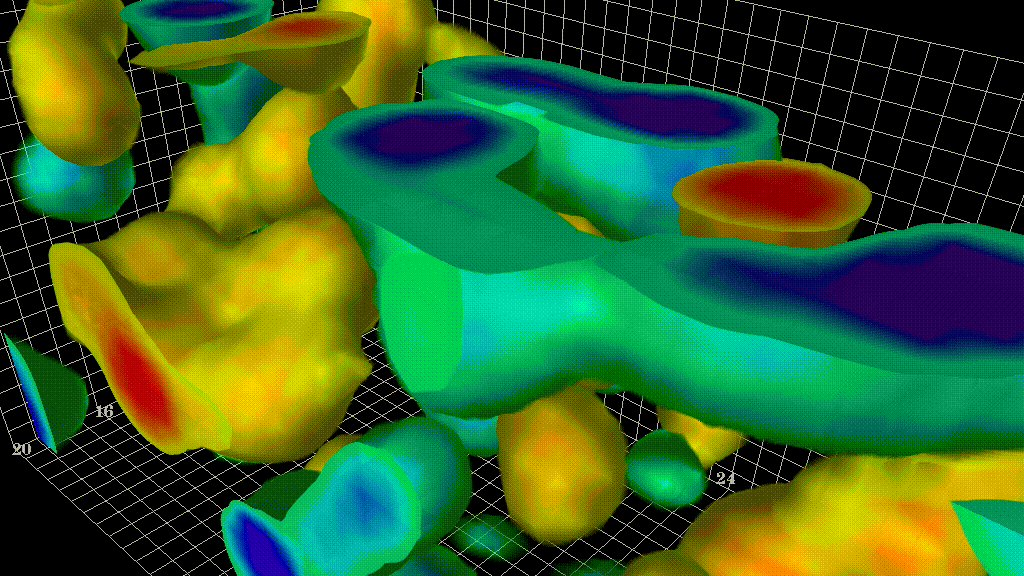}\\
  \includegraphics[width=0.48\textwidth]{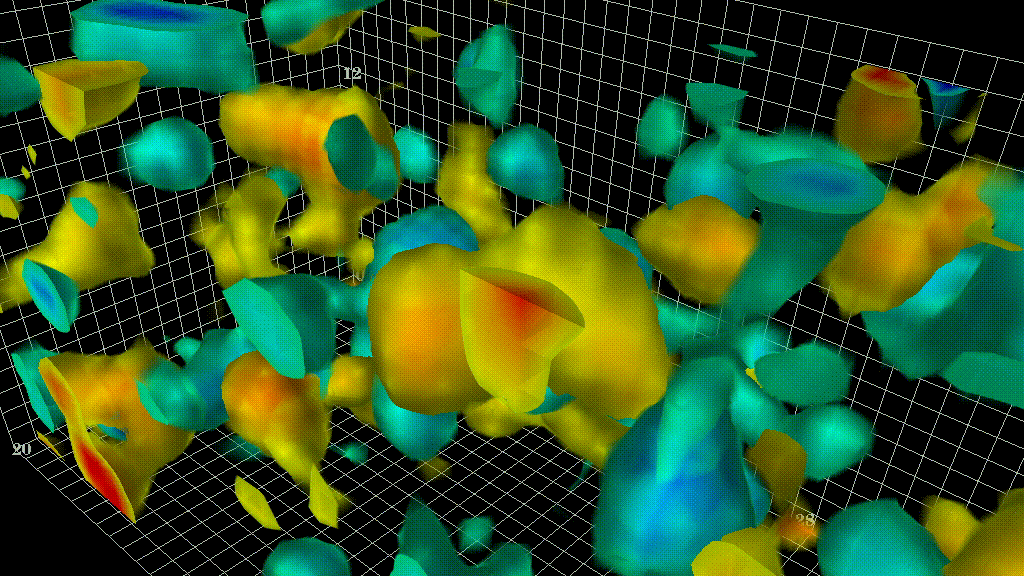}
  \includegraphics[width=0.48\textwidth]{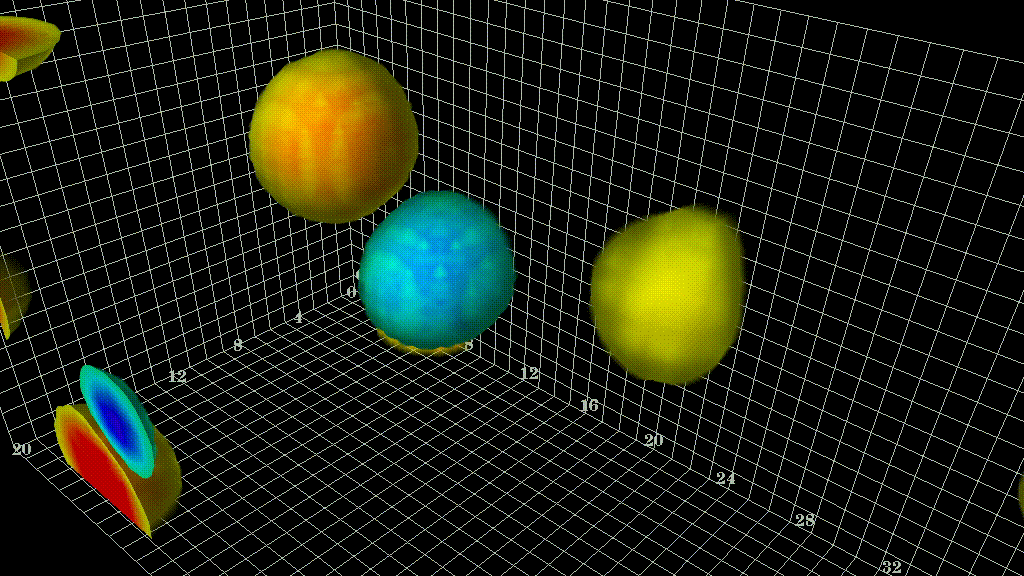}	
\caption{A visualisation of the topological charge density after 10 (left) and (40) sweeps of cooling on a representative configuration, for the untouched (top), vortex-only (middle), and vortex-removed (bottom) fields. Contours of positive topological charge density are plotted in yellow, and negative in blue.}
\label{Fig:cool10q40q}
\end{figure}

Fig.~\ref{Fig:cool10q40q} shows the topological charge density on a representative gauge field configuration. The three rows show the same configuration in the untouched (top), vortex-only (middle) and vortex-removed (bottom) forms. The left column shows the topology after 10 sweeps of cooling. At this point we are able to observe that the topological objects for the untouched and vortex-only fields are similar in size and strength, while the vortex-removed field contains smaller objects with with a reduced charge density at their core. After 40 sweeps of cooling, shown in the right column, the difference is dramatic. The vortex-removed field is now almost entirely devoid of topology. There are some residual objects, associated with imperfections in the SU(3) vortex-identification process which is known to be unable to simultaneously identify all vortex matter~\cite{Cais:2008za}.

\begin{figure}[t]
  \centering
  \includegraphics[width=0.4\textwidth]{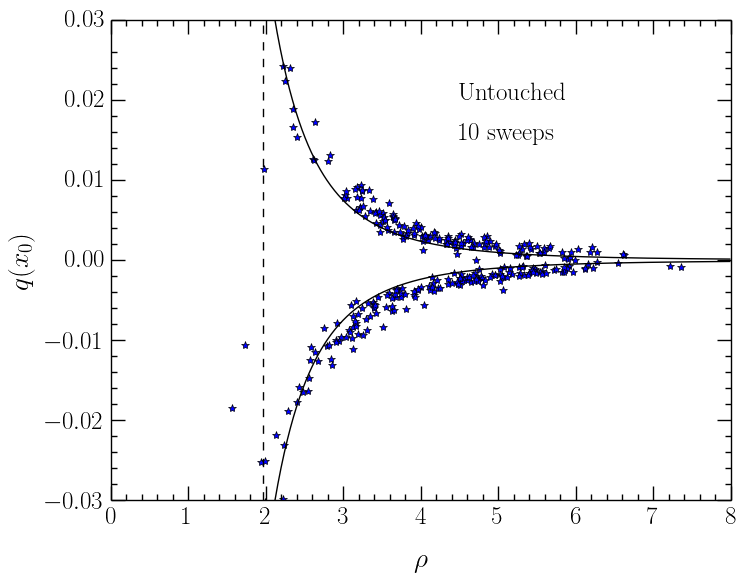}%
  \includegraphics[width=0.4\textwidth]{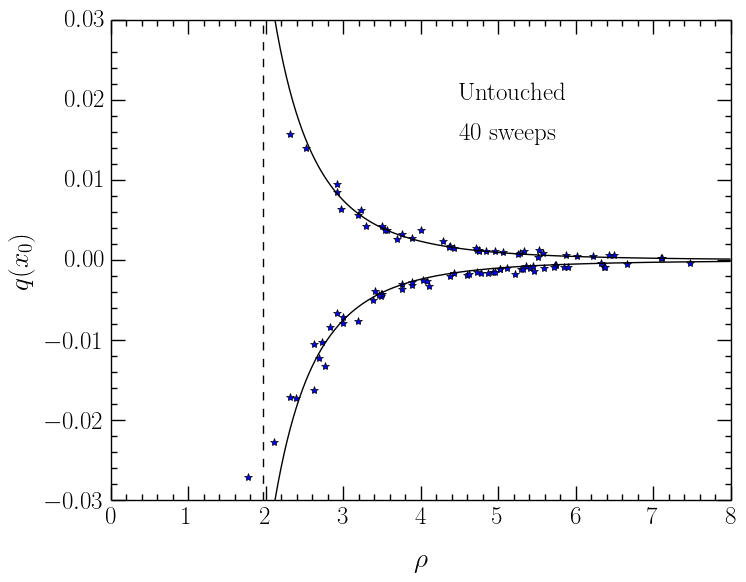} \\
  \includegraphics[width=0.4\textwidth]{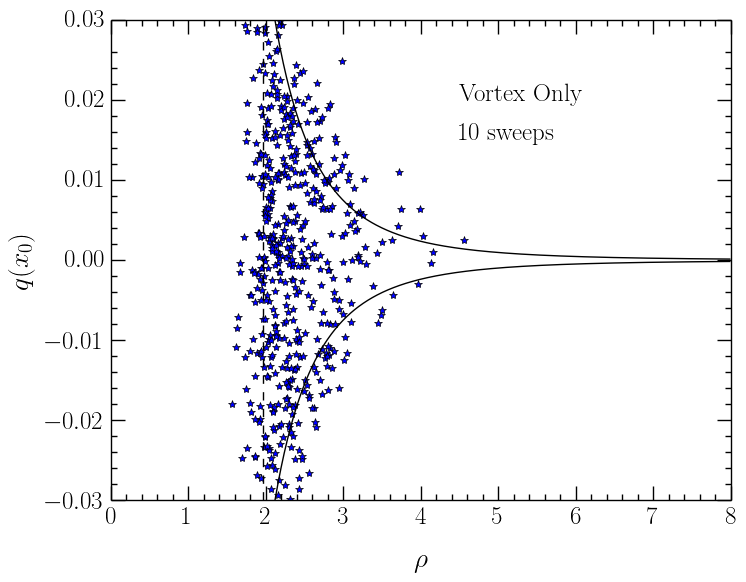}%
  \includegraphics[width=0.4\textwidth]{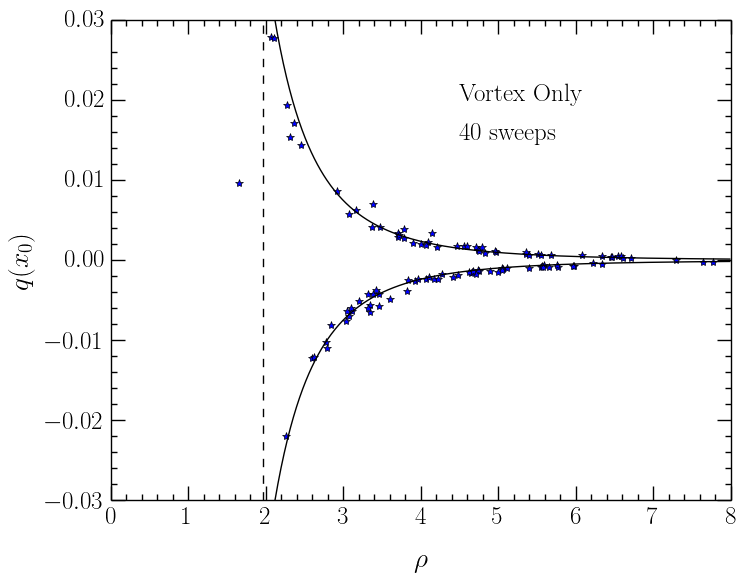} \\
  \includegraphics[width=0.4\textwidth]{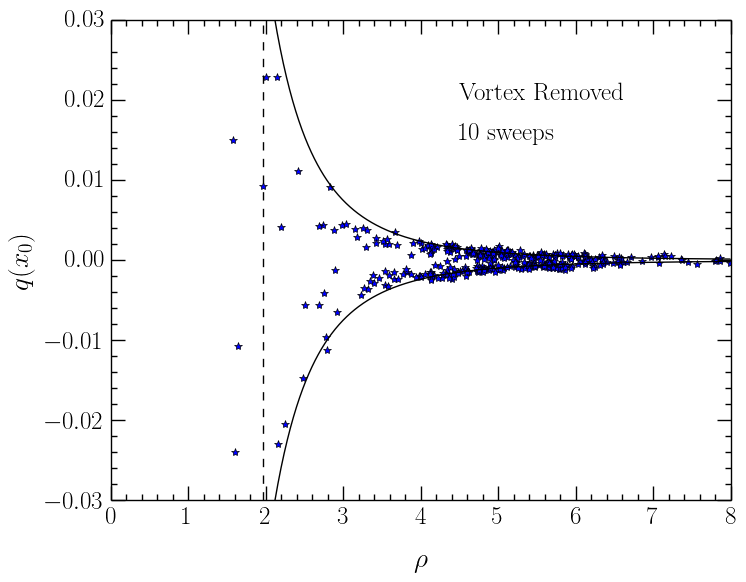}%
  \includegraphics[width=0.4\textwidth]{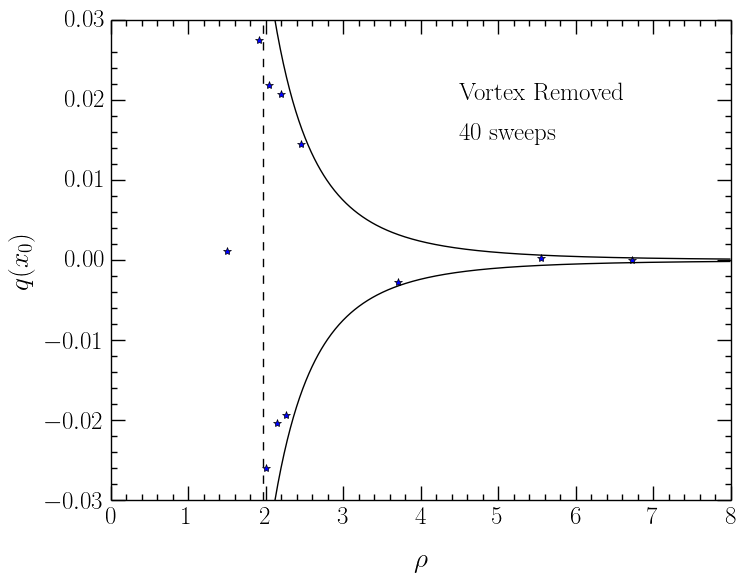}
\caption{The distribution of values of the fitted instanton radius, $\rho$, against the topological charge at the centre, $q(x_{0})$, for a representative configuration. Results are compared to the theoretical relationship between the instanton radius and topological charge at the centre (solid lines), and the dislocation threshold, $1.97a$ (dashed line). Results are shown on a typical configuration for the untouched (top), vortex-only (middle), and vortex-removed (bottom) fields at 10 (left) and 40 (right) sweeps of cooling.}
\label{Fig:cool10RvQ}
\end{figure}

The level of similarity between the untouched and vortex-only fields after 40 sweeps of cooling is striking. While the specific locations of the topological objects are different, the qualitative features of the untouched and vortex-only fields in terms of the strength and size of the topological objects are essentially identical. This leads us to the interesting notion that the thin centre vortices are the seeds of the thick centre vortices, and through the application of cooling we may grow the thin vortices into fully-fledged, instanton-like topological objects. 

This concept is explored further in Fig.~\ref{Fig:cool10RvQ} (reproduced from Ref.~\cite{Trewartha:2015ida}), where the topological objects on a representative configuration are measured by fitting local maxima of the action density to the classical instanton solution, and then comparing with the theoretical relationship connecting the instanton radius with the topological charge at the centre. After 10 sweeps of cooling, all three fields produce distinct distributions, and at this point only the untouched distribution resembles the theoretical relationship. However, after 40 sweeps of cooling, we see that the distribution from the untouched and vortex-only fields are now nearly identical and both closely follow the theoretical curve, while the vortex-removed distribution is almost empty.

\begin{figure}
  \centering
  \includegraphics[width=0.5\textwidth]{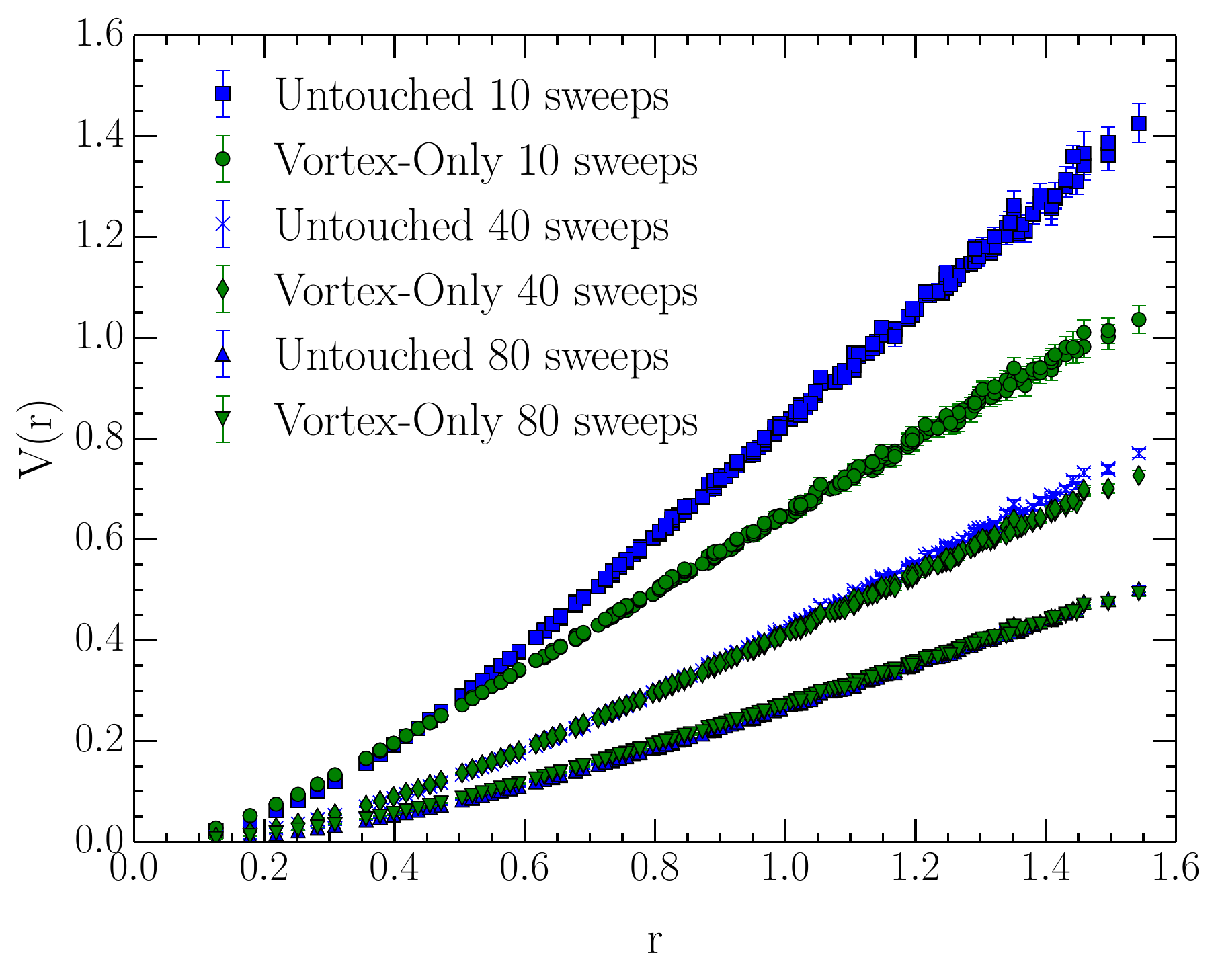}
  \caption{The static quark potential on untouched (blue) and vortex-only (green) configurations after 10, 40, and 80 sweeps of cooling. Note that the 80 sweeps untouched potential is hidden behind the 80 sweeps vortex-only potential.}
  \label{Fig:SQPc}
\end{figure}

Having found evidence of a link between centre vortices and the instanton structure of the vacuum by examining the topological charge density on individual configurations, we now turn our attention to quantities that can be measured at the ensemble level. The connection between the existence of a linear potential, or string tension, and the notion of permanent quark confinement in pure Yang-Mills theory is well established~\cite{'tHooft:1977hy}. Previous results in SU(3) lattice gauge theory show that while the removal of vortices results in the loss of the confining potential, on the vortex-only fields only about two-thirds of the linear potential is recovered~\cite{Bowman:2010zr}. The cause of this deficiency is understood to reside in the vortex identification procedure, which is imperfect. Yet we see in Fig.~\ref{Fig:SQPc} that after only a small amount of cooling the vortex-only fields are able to recreate the string tension of the cooled untouched fields. While the identified centre vortices are inadequate to reproduce the full confinement potential, their evolution under smoothing is sufficient.  It will be interesting to investigate finer lattice spacings under fixed iterations of cooling to learn if this is a lattice cutoff effect or a need to grow long-distance structures from the centre phases.

In addition to confinement, the nontrivial QCD vacuum gives rise to dynamical chiral symmetry breaking, with the resulting non-perturbative quark-gluon interactions yielding the dynamical generation of mass that is responsible for 99\% of the mass of the nucleon. The quark mass function is intimately connected with dynamical chiral symmetry breaking. In a covariant gauge, the lattice quark propagator can be decomposed
into Dirac scalar and vector components as
\begin{equation}
S(p) = \frac{Z(p)}{iq\:\!\!\!\!\!\!/\, + M(p)}\, ,
\end{equation}
with $M(p)$ the non-perturbative mass function and $Z(p)$ containing
all renormalisation information.  The infrared behaviour of $M(p)$
reveals the presence or absence of dynamical mass generation.

\begin{figure}[t]
  \centering
  \includegraphics[width=0.4\textwidth]{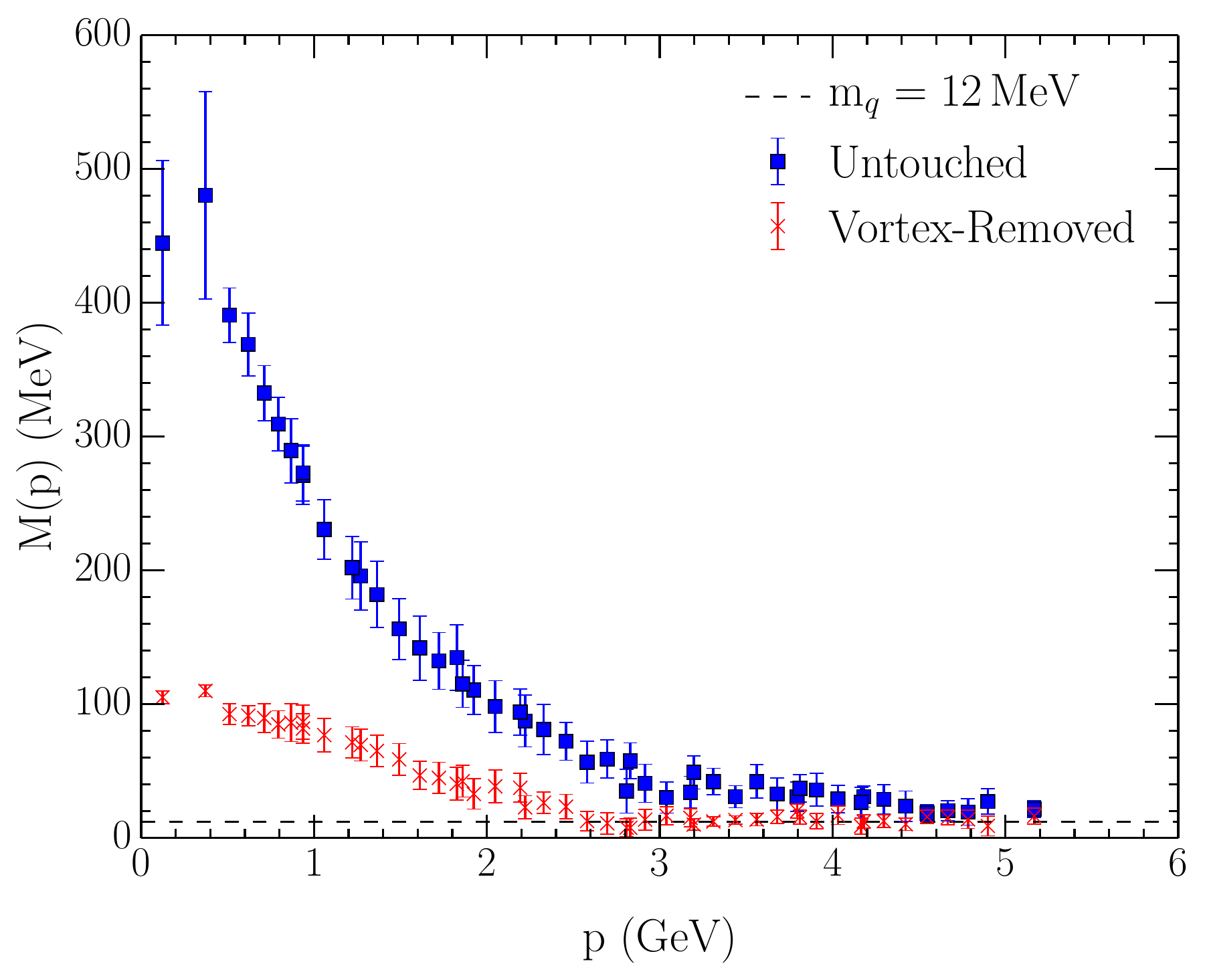}%
  \includegraphics[width=0.4\textwidth]{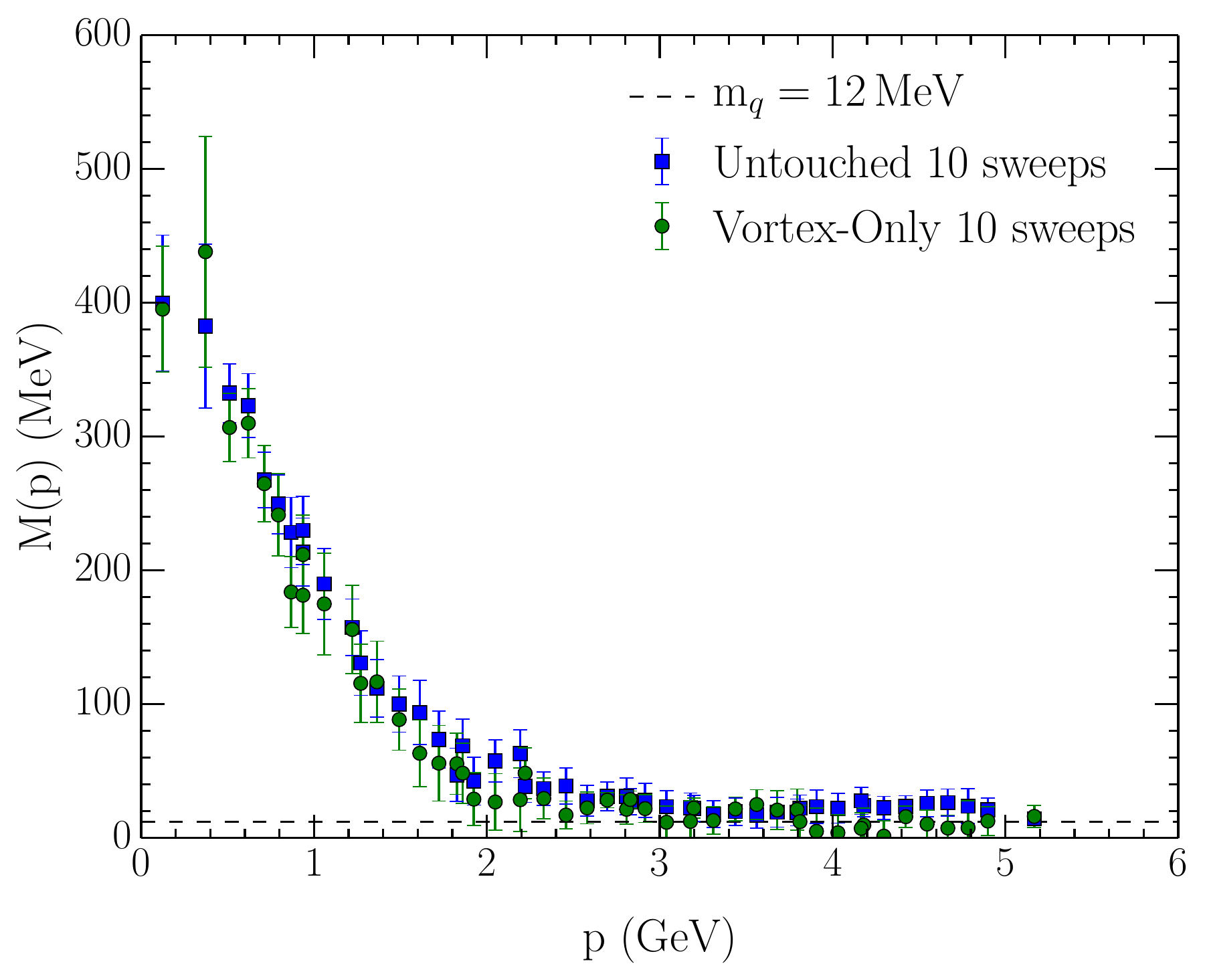}
\caption{ A comparison of the overlap quark mass function in Landau gauge at an input
  bare quark mass of $12$ MeV on the different ensembles. (Left) The
  mass function on the original (untouched) (squares) and
  vortex-removed (crosses) configurations.  Removal of the vortex
  structure from the gauge fields spoils dynamical mass generation and
  thus dynamical chiral symmetry breaking.  (Right) The mass function
  on the original (untouched) (squares) and vortex-only (circles)
  configurations after 10 sweeps of three-loop
  $\mathcal{O}(a^4)$-improved cooling. The cooled vortex-only ensemble
  is able to reproduce the dynamical mass generation present on the
  cooled Monte Carlo gauge fields.}
\label{M00400UTVRcUTcVO}
\end{figure}

The left hand plot of Fig.~\ref{M00400UTVRcUTcVO} compares the quark mass function in Landau gauge on the untouched and vortex-removed ensembles (without any cooling). On the untouched ensemble, the mass function shows strong enhancement in the infrared, indicating the presence of dynamical mass generation. By contrast, dynamical mass generation is largely suppressed upon vortex removal with only a relatively small level of residual infrared enhancement remaining (again, likely to be associated with imperfections in the vortex-removal procedure). The removal of the vortex structure from gauge fields removes dynamical chiral symmetry breaking.

The chiral nature of the overlap operator is critical in being able to `see' the subtle damage caused to the gauge fields through vortex removal. The overlap operator has a smoothness requirement~\cite{Narayanan:1994gw, Hollwieser:2008tq} for the underlying gauge field, and is not well-defined on the rough vortex-only fields consisting solely of centre elements. However, in this instance we are motivated to consider the vortex-only fields after cooling, which addresses the smoothness issue. The right hand plot of Fig.~\ref{M00400UTVRcUTcVO} compares the quark mass function on the untouched and vortex-only ensembles, both after 10 sweeps of cooling. The agreement between the two ensembles is remarkable, with the cooled vortex-only fields able to fully reproduce the dynamical mass generation present on the cooled Monte Carlo gauge fields.

\begin{figure}[t]
  \centering
  \includegraphics[width=0.4\textwidth]{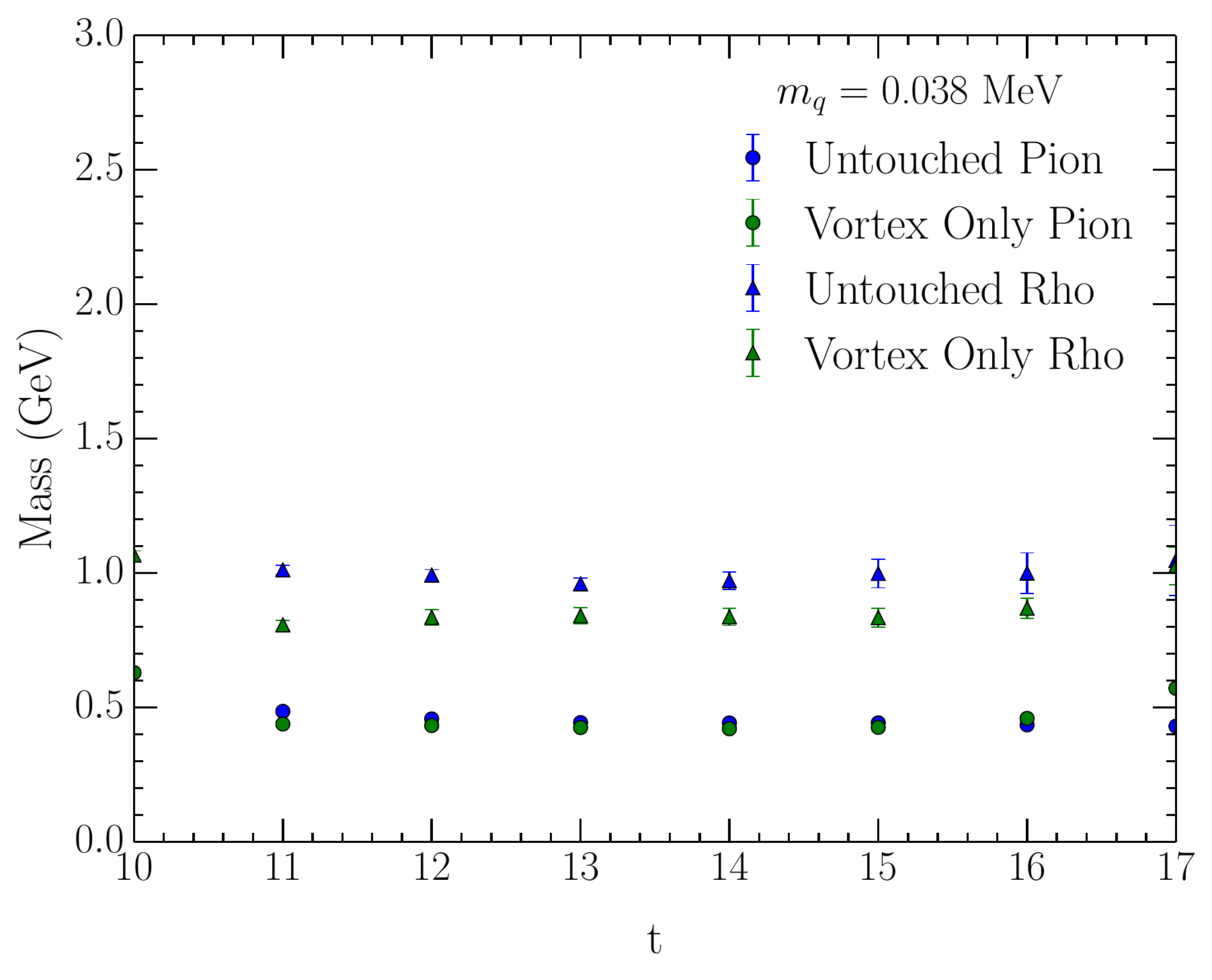}
  \includegraphics[width=0.4\textwidth]{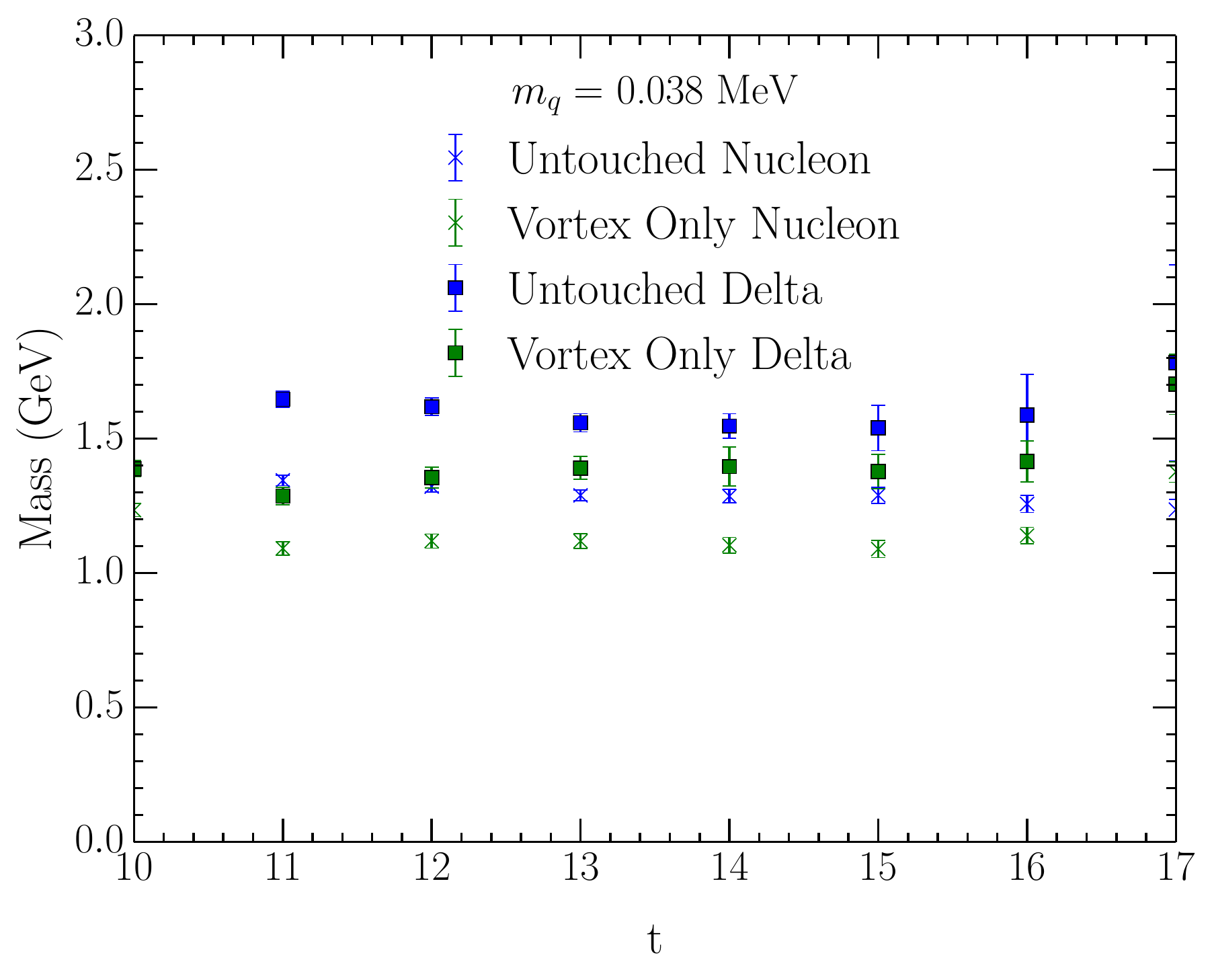}
\caption{The effective masses of the low-lying mesons (left) and baryons (right) considered on the untouched (blue) and (10-sweep cooled) vortex-only (green) ensembles, at a bare quark mass of $38$ MeV.}
\label{Fig:utVOhadrons}
\end{figure}

Of course, dynamical mass generation is fundamentally connected to the hadron mass spectrum. The ability to reproduce the experimentally observed light hadron spectrum is a celebrated feature of lattice QCD. It is interesting to consider what role centre vortices, and hence dynamical chiral symmetry breaking, play in the formation of the hadron spectrum. Noting the potential importance of chiral symmetry, we investigate this question using overlap fermions. Again, the smoothness requirement necessitates using a small amount of cooling on the vortex-only fields. Fig.~\ref{Fig:utVOhadrons} shows the effective mass plots for the light mesons $(\pi,\rho)$ and light baryons $(N,\Delta)$ calculated with overlap fermions on the untouched ensembles (with no cooling) and the vortex-only ensembles (with 10 sweeps of cooling). The cooled vortex-only ensemble is able to recreate the majority of the mass of the light hadrons, as well as the correct ordering. The effective mass for the vortex-only and untouched pion are essentially identical, while for the $\rho, N$ and $\Delta$ the mass on the cooled vortex-only field is roughly 80\% of the value on the original Monte Carlo gauge fields, in accord with the results of Ref.~\cite{Thomas:2014tda}.

The hadron spectrum on the vortex-removed ensemble holds fascinating possibilities. A previous study using Wilson-like fermions~\cite{OMalley:2011aa}, which explicitly break chiral symmetry, showed that the vortex-removed spectrum appeared to be described by weakly-interacting constituent quarks. Here we are using a chiral fermion action, so we are in a position to ask: does the removal of vortices result in the restoration of chiral symmetry? This question leads to the postulation of two different scenarios in the vortex-removed spectrum. At heavy quark masses we should observe a weakly interacting constituent quark regime, where the light hadron masses are simply a result of counting quarks. In this case, the $\pi$ and $\rho$ mesons should be degenerate, as should the $N$ and $\Delta$ baryons. At light quark masses the restoration of chiral symmetry implies that hadron currents related by the appropriate symmetries should become degenerate, while others remain distinct. In particular, the $\pi$ and $\rho$ mesons should have different masses as they are not related by a chiral transformation, while the nucleon and $\Delta$ should have the same mass. Our results for the vortex-removed mass spectrum with overlap fermions are presented in detail elsewhere~\cite{Trewartha:mass}.

In summary, we have examined the topological charge density, the static quark potential, the quark mass function, and the hadron mass spectrum on the untouched, vortex-only and vortex-removed ensembles. In each of these instances we found that starting from the vortex-only fields, which consist only of the centre elements of SU(3), through the application of cooling we are able to reproduce all the salient features of QCD, including confinement and dynamical mass generation. The consistent story provides compelling evidence that centre vortices are indeed the seeds of dynamical chiral symmetry breaking and confinement.
 
\bibliographystyle{JHEP-no-article-title}
\bibliography{reference}

\providecommand{\href}[2]{#2}\begingroup\raggedright\begin{thebibliography}{10}

\bibitem{'tHooft:1977hy}
G.~'t~Hooft, \null {\em Nucl.Phys.} {\bf B138} (1978) 1.

\bibitem{'tHooft:1979uj}
G.~'t~Hooft, \null {\em Nucl.Phys.} {\bf B153} (1979) 141.

\bibitem{OMalley:2011aa}
E.-A. O'Malley et~al., \null {\em Phys.Rev.} {\bf D86} (2012) 054503,
  [\href{http://arxiv.org/abs/1112.2490}{{\tt arXiv:1112.2490}}].

\bibitem{Trewartha:2015ida}
D.~Trewartha et~al., \null {\em Phys. Rev.} {\bf D92} (2015), no.~7 074507,
  [\href{http://arxiv.org/abs/1509.0551}{{\tt arXiv:1509.0551}}].

\bibitem{Trewartha:2015nna}
D.~Trewartha et~al., \null {\em Phys. Lett.} {\bf B747} (2015) 373--377,
  [\href{http://arxiv.org/abs/1502.0675}{{\tt arXiv:1502.0675}}].

\bibitem{Cais:2008za}
A.~O'Cais et~al., \null {\em Phys.Rev.} {\bf D82} (2010) 114512,
  [\href{http://arxiv.org/abs/0807.0264}{{\tt arXiv:0807.0264}}].

\bibitem{Bowman:2010zr}
P.~O. Bowman et~al., \null {\em Phys.Rev.} {\bf D84} (2011) 034501,
  [\href{http://arxiv.org/abs/1010.4624}{{\tt arXiv:1010.4624}}].

\bibitem{Narayanan:1994gw}
R.~Narayanan and H.~Neuberger, \null {\em Nucl.Phys.} {\bf B443} (1995)
  305--385, [\href{http://arxiv.org/abs/hep-th/9411108}{{\tt hep-th/9411108}}].

\bibitem{Hollwieser:2008tq}
R.~H{\" o}llwieser et~al., \null {\em Phys.Rev.} {\bf D78} (2008) 054508,
  [\href{http://arxiv.org/abs/0805.1846}{{\tt arXiv:0805.1846}}].

\bibitem{Thomas:2014tda}
S.~D. Thomas et~al., \null {\em Phys. Rev.} {\bf D92} (2015), no.~9 094515,
  [\href{http://arxiv.org/abs/1410.7105}{{\tt arXiv:1410.7105}}].

\bibitem{Trewartha:mass}
D.~Trewartha, W.~Kamleh, and D.~B. Leinweber, \null {\em in preparation.}

\end{thebibliography}\endgroup

%

\end{document}